\documentclass{IOS-Book-Article}

\usepackage{graphicx}

\usepackage{xspace}

\newcommand{\tem}{\mbox{\textsc{Temanejo}}\xspace}
\newcommand{\ayu}{\mbox{\textsc{Ayudame}}\xspace}

\begin{document}
\begin{frontmatter}                           

\title{\tem - a debugger for task based parallel programming models}

\author[A]{\fnms{Steffen} \snm{Brinkmann}},
\author[A]{\fnms{Jos\'e} \snm{Gracia}},
\author[A]{\fnms{Christoph} \snm{Niethammer}}
and
\author[A]{\fnms{Rainer} \snm{Keller}}

\runningauthor{S. Brinkmann et al.}
\address[A]{High Performance Computing Centre Stuttgart (HLRS), University 
    of Stuttgart, Germany}

\begin{abstract}
We present the program \tem, a debugger for task based parallelisation 
models such as StarSs.
The challenge in debugging StarSs applications lies in the fact that tasks 
are scheduled at runtime, i.e dynamically in accordance to the data 
dependencies between them.
Our tool assists the programmer in the debugging process by visualising the 
task dependency graph and allowing to control the scheduling of tasks.

The toolset consists of the library \ayu which communicates with the 
StarSs runtime on one side and of the debugger \tem on the other 
side which communicates with \ayu . \tem provides a graphical user interface with which the 
application can be analysed and controlled.
\end{abstract}

\begin{keyword}
Debugging\sep HPC \sep task parallelisation\sep StarSs\sep Temanejo
\end{keyword}
\end{frontmatter}

\thispagestyle{empty}
\pagestyle{empty}

\section*{Introduction}

Task based programming models have amplified the landscape of
parallelisation paradigms. In these models, parallelism is not dictated
on the level of operations. The programmer rather indicates data dependencies
between parts of the code which become \emph{tasks} and then are scheduled by an
execution framework at runtime. This way the program can ``react'' on
internal irregularities, e.g. differences in execution time of each task,
and external conditions, such as hybrid architectures. 

One family of these programming models are the StarSs
models~\cite{planas2009,marjanovic2010}.  The ``Ss''
stands for \emph{superscalar} and indicates that these models aim at scaling
on modern supercomputers consisting of millions of processing units. The
``Star'' in the name paraphrases a wildcard which stands for the different
implementations of this model like \emph{CellSs}, \emph{GPUSs},
\emph{GridSs}, \emph{OMPSs} and \emph{SMPSs}~\cite{marjanovic2010} among others. 
They differ in the supported programming languages, 
whether functions (respectively subroutines) or other blocks
of code are parallelised and whether the target platform is, for instance,
a grid or a node of CPUs or GPUs. 

Debugging a task parallel program differs from debugging an otherwise
parallelised program as the parallelisation is determined by data
dependencies instead of explicit scheduling of tasks and synchronisation
between them. Therefore, the development of new debugging tools capable
of assisting the programmer with debugging task parallel programs is
a crucial condition for effectively exploiting the possibilities of
these models.

We present a debugging toolset for task parallel programs consisting
of two components: a library called \ayu
\footnote{Spanish for \emph{help me}}, acting as a thin communication layer and 
enabling the debugger to receive information about the program and
send commands back to it, and the actual debugger \tem \footnote{Spanish for 
\emph{I handle you}} which enables the
user to view the extracted information and issue commands via 
a highly interactive graphical user interface.

For this work, we used the StarSs implementations SMPSs and OMPSs. 
Nevertheless, the basic concepts of
how to debug a task parallel program are the same regardless of the 
specific framework.

In section~\ref{sec:task}, we elaborate on the idea and further implications
of task based parallelism. We present different
strategies and actions a debugger should provide and discuss how these 
features were implemented in the presented debugger \tem in section~\ref{sec:debug}. 
Finally, we summarise the results and give an outlook on future
enhancements to the debugger in section~\ref{sec:sum}.

\section{Task based parallelism}
\label{sec:task}

In contrast to other task based parallel programming models, the focus
in the StarSs family lies on the data
dependencies between parts of the program which are defined as tasks.
In SMPSs, for example, the programmer marks functions or subroutines as 
``potentially parallel'' using a special \mbox{\texttt{\#pragma}} syntax. A 
typical function declaration in \texttt{C} looks like this:

\begin{verbatim}
#pragma css task input(a) inout(b) output(c)
void func(int a, int *b, double *c)
{
    /* function body */
}
\end{verbatim}

\texttt{css}\footnote{\texttt{css} 
stands for \texttt{cell super scalar} and is a historical remnant of the first 
versions of SMPSs.} is the identifier of the pragma, the keyword \texttt{task}
tells the precompiler that the following function declaration is to be
treated as an SMPSs task, and the keywords \texttt{input}, \texttt{inout} and
\texttt{output} specify the data dependencies for this task.

These pragmas are read by a compiler wrapper (\texttt{smpss-cc}) which 
embeds the program code into a runtime framework. This framework will initialise 
a previously defined number of threads and start the application.
One of the threads will be the \emph{master thread} which creates all tasks
and takes care of proper initialisation and finalisation of the program.
Moreover the master thread will execute all code outside of tasks. The tasks
will be executed by the other threads, called \emph{worker threads}

During runtime the \texttt{SMPSs} framework will assign all calls to functions marked by 
pragmas to a thread for execution depending on the specified dependencies.

\begin{figure}[!t]
\centering
\includegraphics[width=.5\columnwidth]{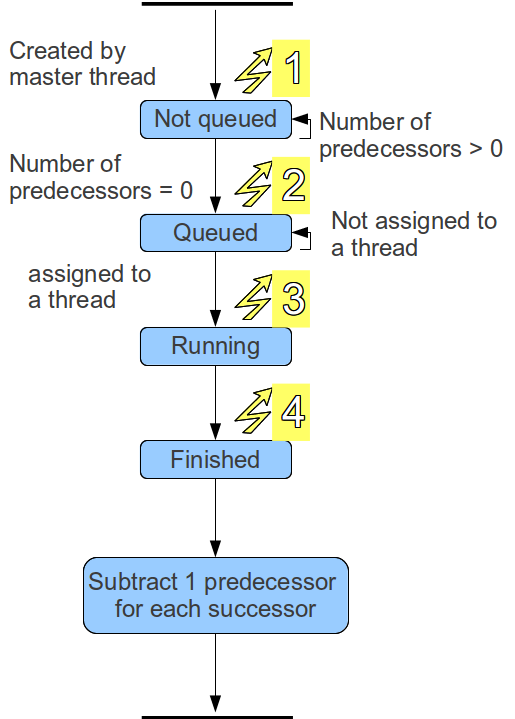}
\caption{\textbf{State transition diagram of a task}. 
The task is always in one of the states \emph{``not queued''}, 
\emph{``queued''}, \emph{``running''} or \emph{``finished''}. The processing
of the successors is not counted as an own state. The transition
to a state is triggering an event indicated by the numbers.}
\label{fig:task}
\end{figure}

To achieve this, a task must be in one of four states (see figure~\ref{fig:task}):
\emph{``not queued''}, \emph{``queued''}, \emph{``running''} or
\emph{``finished''}. When tasks are created they are \emph{not queued} in case 
they have pending dependencies, i.e., they read (\emph{input}) from a memory 
address which has been declared \emph{output} (or \emph{inout}) by another task
at an earlier time. Otherwise, the task is \emph{queued} directly after creation. 

When all tasks on which a task depends on are finished, the task is queued. 
A queued task can be run at any time by an idling thread. When a task is executed
by a thread, its state changes to \emph{running}. After finishing the taks, its
state is \emph{finished}.
 
The transition between the states is a useful information for the programmer
(see section~\ref{sec:vis}). Furthermore, the state transition diagram 
(figure~\ref{fig:task}) indicates
that there is more than one way to block or prioritise a task. In the case of
blocking, one can avoid for a task to be queued or avoid that it is ever dequeued
for execution by a thread. While a task is blocked, one can let the other
threads continue executing tasks or interrupt the program execution entirely.
Other possibilities for debugging include stopping only a certain function or adding
artificial dependencies in order to force a certain order of subtree 
execution.

Some of these possibilities are exploited by the debugger \tem. How this
was done is explained in the following. An outlook on future features
is given in section~\ref{sec:sum}.

\section{The debugger \tem}
\label{sec:debug}

In this section we will investigate the different requirements regarding a 
debugging tool for task based problems and how the debugger \tem fulfils them.
A rather technical introduction to \tem is given in the next section.

For the programmer it is necessary to see (section~\ref{sec:vis}) and 
be able to manipulate (sections~\ref{sec:block}, \ref{sec:stop} and \ref{sec:prio})
the dependencies which lead to the conditions for executing tasks and thus 
alter the way the runtime will step through the tasks. As
an additional information the duration of each task can be displayed 
(section~\ref{sec:dur}).

In contrast to other parallelisation strategies, when using task based models
the programmer needn't (and cannot) know a priori when a certain task
is executed.  At compile time only the conditions for running a task are
known.  Therefore, classical debuggers which step through lines 
of code sequentially are not well suited for debugging task parallel programs.
They form a part of the debugging process (see section~\ref{sec:gdb}) but need to be 
complemented by other tools. 

\subsection{Technical details}
\label{sec:tech}

\tem is written in python and requires python version 2.6 
or newer. Furthermore, it imports several modules, the most important being
Networkx~\cite{networkx} for the graph data structure and
pygtk~\cite{pygtk} for the visualisation and the graphical user interface.

The debugger \tem is tightly linked to the small \mbox{library} \ayu.
This \mbox{library} serves as a thin communication layer between the runtime 
and \tem and has to be preloaded to the application.
The SMPSs and OMPSs framework (thus indirectly the application) are instrumented
with calls to an event handler which passes the information to the debugger
as a set of 8 integer values containing an event id, the task id and, depending on
the event, one or more of the following: the task id of a dependency, the memory 
address of a dependency, the id of the function assigned to the task, a thread id and/or
a timestamp. 
This is done at certain crucial points in the program execution, for
instance initialisation, task generation, task execution etc., 
the event function is called and will react accordingly. In most
cases, it will simply pass the data (task id, thread id, dependencies
and so forth) to the attached socket client (in this case 
\tem ).

\tem uses this information to build an internal data 
structure representing the dependency graph.
Each task is a node of this graph and each dependency an edge. 
Nodes and edges have properties such as the task status and the
memory address of the dependency, respectively. 

The information is displayed by \tem at the time it is received and the user can access
the debugging features instantly by pressing buttons or right-clicking on nodes.
There are four distinct properties available for displaying information:
the node colour, the node-margin colour, the node
shape and the edge colour. This way one can produce different representations
of the same graph depending on what information is selected for 
visualisation at a given debugging session.

In the current version (0.5) \tem relies on connecting to the \ayu 
library as a socket client when starting up. This means that either the parallel
application has to run when \tem is started or \tem will start the application
at the beginning of the debugging session.

\subsection{Visualisation of the dependency graph}
\label{sec:vis}

\begin{figure}[!t]
\centering
\includegraphics[width=.85\columnwidth]{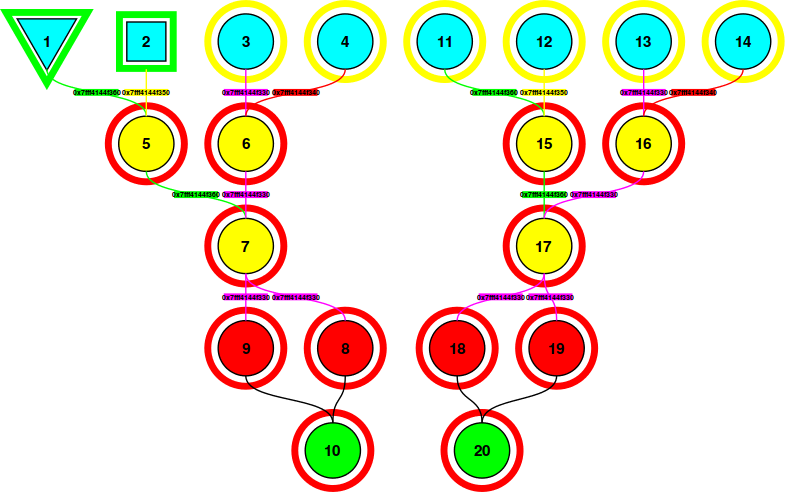}
\caption{\textbf{A simple dependency graph} with two independent
subtrees consisting of ten tasks each visualised by \tem . 
Eight independent tasks are drawn in blue, two of them scheduled 
(green margin) and six of them queued (yellow margin). 
Three tasks with input and output dependencies 
are shown in yellow, four reduction tasks in red and two tasks with only
one input dependency in green. The red margin indicates that a 
task has unfulfilled dependencies and can therefore not be queued yet.
The node shapes (triangle and box) denote two different worker threads,
the circle shape indicates that the task has not been assigned to
a thread yet. The text labels and colours of the edges indicate the 
memory addresses of the dependencies.}
\label{fig:depgraph}
\end{figure}

The first step of data-flow oriented debugging is to visualise
the dependency graph. Often this already helps the 
programmer to spot errors or other shortcomings of his program. An 
example of such a dependency graph as displayed by \tem is shown in 
figure~\ref{fig:depgraph}.

It is important to generate and show the dependency graph while the 
program is running (``online'') rather than just collecting the 
dependency information for later review (``offline''). 
Firstly the tasks are generated dynamically during runtime
hence sometimes the solution to a problem lies in the order in which
tasks are generated. Secondly viewing the graph after running the parallel
program makes the graph simply the \emph{result} of a program rather 
than a \emph{tool} to debug it. 
Therefore, passing information during 
runtime from the program to the debugger is essential. 
For SMPSs and OMPSs, this is achieved by linking to the library 
\ayu .

\subsection{Controlling Task execution}
\label{sec:taskcontrol}

The second requirement for debugging is for the programmer to be able to interact
with the running application. This means that the programmer can block a task,
stop the program execution at some point, prioritise a task and attach a \texttt{gdb}
session to the application. How these features are achieved by \tem, is described in 
the following.

\subsubsection{Blocking single tasks}
\label{sec:block}

In order to block a specific task, the threads have to be kept from
scheduling it for execution. At the same time it must be possible to unblock
a task again. We achieve this by building a "\emph{to-be-blocked}" list
in the runtime from the user input to \tem . Whenever a thread wants to execute a task, it checks
whether the task is in that list and if so he skips it and searches the queues for another one.

\subsubsection{Stop Execution when reaching a task}
\label{sec:stop}

Stopping the program execution at a certain point means to avoid the
dequeueing of tasks for any thread. This is implemented by a global flag
which is checked by an idling thread \emph{before} looking for a new task in the queues. If the flag is set,
the thread is sent back to idle.

\subsubsection{Prioritising tasks}
\label{sec:prio}

In order to prioritise a task, we make use of SPMSs' ``critical'' queue which is
also accessible by adding the keyword \texttt{highpriority} to the pragma line
before the function definition. With a right-click on the node in \tem the user
can prioritise and \emph{de}prioritise any task before it queued. At the present
stage it is not possible to ``requeue'' a task, i.e. to take a task from one queue
and put it into another. In future versions this will be possible which will
enable the user to prioritise a task which is already queued in order to execute
a specific task as early as possible.

\subsubsection{Attaching \texttt{gdb}}
\label{sec:gdb}

\tem is equipped with the possibility of launching \texttt{gdb} at any point
of the debugging session. By default the terminal user interface of \texttt{gdb}
starts and lets the user debug any part of the application in the usual line
based way.

Another way to launch \texttt{gdb} is by specifying a function in \tem at which
\texttt{gdb} should set a breakpoint. Again, a \texttt{gdb} process is attached
to the application process, a breakpoint is set at the specified function and
the program execution is continued, i.e. the control is given back to \tem. When
the program execution reaches the function, \texttt{gdb} takes control and 
the user can debug the function with \texttt{gdb}.

\subsection{Task duration analysis}
\label{sec:dur}

In order to enable the StarSs framework to balance each thread's load, the task
must neither be to small nor too big. For SMPSs the optimal duration of a task in order
to keep all threads busy while avoiding too much overhead wasted in generating,
queueing, scheduling and cleaning up tasks, was shown to be of the order of hundreds
of milliseconds. An application with tasks which run less than $\approx 100\mu{}s$ will
suffer a huge overhead in the sense that the master thread will be busy managing
the tasks while the worker threads will spend most of their time waiting for tasks.
On the other hand, when some tasks are to large, the application runs into a load balancing
problem because the worker thread with the large task will again keep other worker
threads waiting.

Therefore, it is important for the programmer to know how long the tasks last. In \tem
the task durations can be shown in colour code for finished tasks. As the communication
with the debugger alters the collection of time information, this can only be a rough
guide and does not replace a profiling tool. Nevertheless, the ability of displaying
timing data in the dependency graph effortlessly proved to be a powerful tool
for programmers.

\section{Summary and outlook}
\label{sec:sum}

We presented \tem , a debugger for task parallel applications. The StarSs implementations
SMPSs and OMPSs in conjunction with the library \ayu enables the
programmer to write task parallel codes and debug them with \tem .

\tem is in constant development process but already provides powerful features
which help to debug and enhance task parallel applications. Foremost it shows 
the dependency graph while it is built up and processed during the runtime of
the application. The node colour, the node margin colour, the node shape and the 
edge colour are four properties which can assigned to display one of the following
information: the function which the task executes, the thread which executes the task,
the status of the task, the duration of each task (all of the before mentioned
for node properties) and the dependency address (for the edge colour).

The presented version \tem 0.5 already is a strong tool for the programmer
to debug applications parallelised with SMPSs and OMPSs. Nevertheless, many 
features are desirable in order to make the debugging even more convenient.

In future versions, we will implement more possibilities to analyse the dependency
graph. Interesting information about the graph are longest and shortest paths,
prioritisation recommendations based on this information, and displaying (or greying
out) subgraphs. This will give the developer deeper insights into dependency graphs
which become quickly very complex in real-life applications.

Furthermore, we will enhance the control of the application giving better access to queues.
This will make it possible to block, prioritise or even run tasks at any time.
With the intelligent automatised use of these features it will also be possible to
run tasks in a predefined order, e.g. in order of generation, the same order of
a previous run or any arbitrary order defined in a configuration file.

Another important tool will be to change the number of threads at any time of 
the program execution. By that it will be possible to serialise the application
at any point and get back to parallel execution afterwards.

Also, we plan to enable the user to launch any debugger instead of just \texttt{gdb}.

Finally, we point out that \tem is, in spite of its stable operation and richness of
feature, still in the process of active development. We encourage the reader to use it
and send feedback of any kind in order to enhance this powerful and unique debugger
in the most useful way.

\section*{Acknowledgements}

This work was supported by the European Community's Seventh
Framework Programme [FP7-INFRASTRUCTURES-2010-2] project TEXT under
grant agreement number 261580. The authors would like to thank Dr. C.~Glass
for his valuable help to enhance this article.

\bibliographystyle{unsrt}
\bibliography{Ayudame-2011-ParCo}

\end{document}